# Detection and Prevention of New Attacks for ID-based Authentication Protocols

JINYONG CHEN

School of Computer Science, University College Dublin, Dublin, Ireland

REINER DOJEN

Department of Electronic & Computer Engineering, University of Limerick, Limerick, Ireland

ANCA JURCUT

School of Computer Science, University College Dublin, Dublin, Ireland

The rapid development of information and network technologies motivates the emergence of various new computing paradigms, such as distributed computing, and edge computing. This also enables more and more network enterprises to provide multiple different services simultaneously. To ensure these services can conveniently be accessed only by authorized users, many password and smart card-based authentication schemes for multi-server architecture have been proposed. In this paper, we review several dynamic ID-based password authentication schemes for multi-server environments. New attacks against four of these schemes are presented, demonstrating that an adversary can impersonate either legitimate or fictitious users. The impact of these attacks is the failure to achieve the main security requirement: authentication. Thus, the security of the analyzed schemes is proven to be compromised. We analyze these four dynamic ID-based schemes and discuss the reasons for the success of the new attacks. Additionally, we propose a new set of design guidelines to prevent such exploitable weaknesses on dynamic ID-based authentication protocols. Finally, we apply the proposed guidelines to the analyzed protocols and demonstrate that violation of these guidelines leads to insecure protocols.

## 1 INTRODUCTION

The security of electronic networks and information systems is a critical issue for the use of communication technologies, particularly so in e-commerce. Mobile and fixed networks are nowadays trusted with highly sensitive information. However, a public network such as the Internet should be treated as an insecure network, where a malicious adversary can eavesdrop, intercept and modify messages. Thus, cryptographic protocols are required to ensure the security of both the infrastructure itself and the information that runs through it [1, 2]. Basic security protocols allow agents to authenticate each other, to establish fresh session keys for confidential communication and to ensure the authenticity of data and services. More advanced services like non-repudiation, fairness, electronic payment and contract signing can be built using these basic security protocols. The design of such security protocols should be robust enough to resist attacks [3], [35].

While password-based authentication or other single factor authentication schemes are comparatively simple to implement, they offer limited security. On the other hand, multi-factor authentication schemes are more complex to implement, but they provide superior security [4]. Multi-factor authentication schemes require participants to present at least two separate authentication factors from the following categories:

- *Knowledge Factors:* Participants need to prove knowledge of a secret, such as a password, passphrase or personal identification number (PIN).
- *Possession Factors:* Participants need to prove possession of a security token, such as a smart card, dongle or wireless tag.
- *Inherence Factors:* Factors associated with the participant, usually based on biometric techniques such as fingerprint scanning, retina scanning and voice or face recognition.

Many multi-factor authentication schemes that are based on different authentication factors have been proposed [4, 5, 6, 7]. Further, many applications require protection of the users' anonymity. This can be achieved by replacing the true identity of users in any transmission with a dynamic identity [8, 9]. Additionally, as more and more services are offered online, users are likely to access resources on multiple servers. If all servers act independently of each other, then users need to register for each server and they are provided each time with a different set of credentials (identities, passwords, possession factors). Consequently, users need to manage their corresponding credentials. Thus, in multi-server environments, it is desirable that users perform a single registration and use a single set of credentials to access individual servers.

**Contribution of this paper.** In this paper, we reveal hitherto unknown weaknesses in the authentication schemes of Liao and Wang [11], Hsiang and Shih [12], Lee et al. [13] and Li et al. [10] and we present new attacks that exploit these weaknesses. We discuss why these new attacks succeed and we propose a new set of design guidelines to prevent such exploitable weaknesses on dynamic ID-based authentication protocols.

The remainder of this paper is organized as follows: Section 2 discusses related works and review several dynamic ID-based authentication schemes. In Section 3 we reveal hitherto unknown weaknesses in the authentication schemes of Liao and Wang, Hsiang and Shih, Lee et al., and Li et al.; and we discuss the new attacks that exploit these weaknesses. In Section 4 we analyze the security weaknesses leading to the presented attacks; introduce a new set of design guidelines to prevent such exploitable weaknesses on dynamic ID-based authentication protocols and apply the proposed guidelines to the analyzed protocols. Section 5 concludes this paper.

## 2  RELATED WORK

In 1981, Lamport [14] formulated a remote password authentication scheme to be used in an insecure communication environment. However, as in this scheme servers maintain authentication tokens in a verification table, it cannot resist an interpolation attack if an adversary breaks into any server [13]. T. Hwang et al. [15] proposed a non-interactive password authentication scheme using smart cards that operates without storing verification tables, but it does not scale well, as it suffers from a complicated registration phase and complex password change procedure. M.S. Hwang and Li [16] proposed a public key-based remote user authentication scheme using smart cards that also operates without maintaining verification tables. However, the M.S. Hwang and Li scheme is vulnerable to attack, where a legitimate, but malicious, user can impersonate other users [17]. Subsequently, many smart card-based authentication schemes using one-way hash functions have been published to improve computational efficiency [18, 19, 20, 21, 22, 23, 24].

All these schemes are designed for single server environments. In 2000, Lee and Chang [25] proposed a user identification and key distribution scheme for multi-server environments. Numerous smart card authentication schemes for multi-server environments have since been published: In 2001 Li et al. [26] proposed a remote user authentication scheme using neural networks that provides a single registration to enable access to different servers. However, Juang [27] observed that the Li et al. protocol provides only unilateral authentication and does not provide a session key agreement mechanism. Juang also proposed an efficient multi-server authentication and key agreement scheme based on a hash function. Chang and Lee [28] demonstrated that the Juang scheme was vulnerable to offline dictionary attacks and online guessing attacks if the secret value in the smart card was extracted. To amend the weakness and reduce the computational cost, they published an improved remote user authentication scheme. Tsaur et al. [29] revealed that the Chang and Lee protocol was vulnerable to insider attacks, spoofing attacks and registration centre spoofing attacks. They proposed an authentication scheme using the RSA cryptosystem and Lagrange interpolation polynomials for a multi-server environment. However, due to its high communication requirements and computational costs the Tsaur et al. scheme is not suitable for low-power devices such as smart cards [10]. Furthermore, these authentication schemes for multi-server environments do not protect the anonymity of users. They are based on static IDs, where users employ the same ID to access the remote server in subsequent connections. In static ID-based schemes, it is easy for an adversary to trace the source of communications to identify remote users [13]. Consequently, dynamic IDs that change with every connection request should be used to protect the anonymity of users.

In 2009, Liao and Wang [11] proposed a dynamic ID-based remote user authentication scheme to achieve user anonymity. This scheme uses hash functions to provide a robust mutual authentication mechanism for a multi-server environment. Further, Liao and Wang claimed that the scheme can achieve mutual authentication, provide two-factor security and guarantee users anonymity.

Hsiang and Shih [12] show that the Liao and Wang scheme is vulnerable to several attacks, including insider attacks, stolen smart card attacks, server spoofing and registration centre spoofing attacks. For example, if an adversary obtains a smart card by malicious means, the stored values can be extracted [30]. This allows the adversary to impersonate the legitimate owner of the smart card without knowing the password. Further, this scheme features poor reparability and fails to provide mutual authentication (due to calculation errors). To remedy these weaknesses, Hsiang and Shih proposed an improved version.

Lee et al. [13] demonstrated that the Hsiang and Shih scheme is vulnerable to several attacks, including masquerading attack and server spoofing attack. For example, any legal user $U_i$ can impersonate any other legal user $U_j$ to access remote servers without knowing the secret password of $U_j$. Further, this scheme also features poor reparability and fails to provide mutual authentication (due to calculation errors). To overcome the identified weaknesses, Lee et al. proposed a new dynamic ID-based authentication scheme.

Subsequently, Li et al. [10] revealed that the Lee et al. scheme does not provide authentication and cannot resist forgery attacks and server spoofing attacks. Furthermore, the personal password change phase is inefficient, as it requires communicating with the registration centre via a secure channel. Finally, Li et al. proposed an improved authentication and key agreement scheme to overcome the revealed weaknesses.

In the next section, we will demonstrate that the authentication schemes of Liao and Wang, Hsiang and Shih, Lee et al. and Li et al. contain further hitherto unknown critical flaws that can be exploited by an adversary.

## 3   NEW ATTACKS ON DYNAMIC ID-BASED PASSWORD AUTHENTICATION SCHEMES

The dynamic ID-based password authentication schemes of Liao and Wang [11], Hsiang and Shih [12], Lee et al. [13] and Li et al. [10] work on the basis that it is not possible to simultaneously guess or compute the identity $ID_i$ and password $PW_i$ of user $U_i$ in polynomial time. Further, they claim that their schemes are secure even if an attacker can extract values stored in smart cards as discussed in [30]. In essence, they attempt to achieve mutual authentication by principals demonstrating their ability to generate secret values.

However, we show that this approach is flawed and we reveal hitherto unknown weaknesses in these four protocols that can be exploited by an adversary. As a result of these weaknesses, the adversary can impersonate other users without knowing either their $ID_i$ or $PW_i$. As authentication is the fundamental goal of these schemes, they do not achieve their primary objective and, therefore, should be considered insecure.

Our attacks on these schemes require that the adversary can extract stored values from a smart card. Further, the adversary in these attacks is either a legitimate user extracting values from his/her own smart card or an arbitrary entity who extracts values from a genuine card obtained through theft or other malicious means. While smart cards are widely considered to be tamper resistant, many authors of authentication schemes believe such assumption may be difficult in practice [needs a ref]. Also, the authors of the authentication schemes discussed in this section claim that their schemes are secure under the assumption that adversaries can extract secret information from the smart card [10, 11, 12, 13]. Table 1 summarizes the notations used throughout this paper.

Table 1: Definition of notations used in this paper

| Notation | Definitions |
| --- | --- |
| $U_i$ | Arbitrary user |
| $S_j$ | Arbitrary server |
| $RC$ | Registration centre |
| $SC$ | Smart Card Reader |
| $ID_i$ | Unique identification of user $U_i$ |
| $W_i$ | Unique password of user $U_i$ |
| $SID_j$ | Unique identification of server $S_j$ |
| $DID_i$ | Dynamic identity of user $U_i$ |
| $h(.)$ | A one-way hash function |
| $K_{rc}$ | The master secret key of registration centre |
| $N_b$ | Nonce generated by user $U_i$ at registration phase |
| $N_i$ | Nonce generated by $U_i$'s smart card |
| $N_{rc}, N_r$ | Secret numbers generated by registration centre |
| $\oplus$ | The exclusive-or operation |
| $\|\|$ | The concatenation operation |
| $X^*$ | Distinguishes received values from stored values |
| $X'$ | Distinguishes calculated values from stored values |
| $X^A$ | Value forged by attacker |

### 3.1 New Attack against Liao and Wang Scheme

In this section, we reveal a new attack against the scheme that allows an adversary to impersonate a fictitious user to access a server $S_j$.

### 3.1.1 Liao and Wang Scheme.

The Liao and Wang scheme has four phases: Registration, Login, Authentication & Verification, and Password Change. Initially, the registration centre RC creates Nrc, Krc and distributes these to all registered servers. In the registration phase user Ui submits personal password PWi and identity IDi to RC, who returns a smart card containing (Vi, Bi, Hi, h(.), Nrc), where Ti = h(IDi || Krc), Vi = Ti ⊕ h(IDi || PWi), Bi = h(PWi) ⊕ h(Krc), Hi = h(Ti). Login and Authentication & Verification phases authenticate the user to the server with the following steps, where DIDi = h(PWi) ⊕ h(Ti || Nrc || Ni); Pij = Ti ⊕ h(Nrc ||Ni || SIDj); Qi = h(Bi || Nrc || Ni); SA = h(Bi || Ni || Nrc || SIDj); UA = h(Bi || Nj || Nrc || SIDj); and SK = h(Bi || Ni || Nj || Nrc || SIDj):

1. Login Request             U->S: (DIDi, Pij, Qi, Ni),
2. Server Acknowledge      S->U: (SA, Nj)
3. User Acknowledge        U->S: (UA)
4. Message Exchange Using Session Key $SK$

The password change phase allows users to change their personal password frequently.

### 3.1.2 New Attack against Liao and Wang Scheme to Impersonate Fictitious Users.

In addition to the known weaknesses discussed in section 2, we show that the Liao and Wang scheme suffers from a hitherto unknown weakness that can be exploited by impersonating a fictitious user to server $Sj$. The server is not able to detect that an illegal request has been made and will accept the login request as a legitimate.

In this scenario, the adversary is a legitimate user $Ua$ with genuine access credentials from the registration centre RC. The adversary retrieves the secret tokens (Va, Ba, Ha, h(.), Nrc) from his/her own smart card and obtains $h(Krc) = Ba \oplus h(PWa)$. Then the attack is mounted as follows:

A1. The adversary generates a random nonce $NPWi$ to replace PWa. Another nonce NTi is generated to replace the value Ta. The adversary also computes the fictitious value $Bi^A = h(N_{PWi}) \oplus h(Krc)$

A2. As per the protocol description, the adversary generates nonce Ni and uses the obtained values to compute:
DID = h(PWi) ⊕ h(Ti || Nrc || Ni) = h(N_PWi) ⊕ h(N_Ti || Nrc || Ni)
Pij = Ti ⊕ h(Nrc || Ni ||SIDj) = N_Ti ⊕ h(Nrc ||Ni || SIDj)
Qi = h(Bi || Nrc|| Ni) = h(Bi^A || Nrc || Ni)
Subsequently, the adversary sends counterfeit login request $(DIDi, Pij, Qi,)$ to service provider $Sj$.

A3. Sj attempts to validate the request by computing:
Ti = Pij ⊕ h(Nrc ||Ni || SIDj) = N_Ti
The server Sj is unable to detect that $Ti = N_{Ti}$ rather than $Ti$ = h(IDi || Krc) and Sj calculates:
H(PWi) = DIDi' ⊕ h(Ti || Nrc || Ni) = h(N_PWi); Bi = h(PWi) ⊕ h(Krc) = h(N_PWi) ⊕ h(Krc) = Bi^A; and
Qi' = h(Bi || Nrc || Ni) = h(Bi^A || Nrc || Ni)
The calculated value Qi' is compared with the received Qi* from the login request. As these match, $Sj$ accepts the login request and computes SA = h(Bi^A || Nrc || Ni || SIDj).

A4. The adversary accepts the server acknowledgement message, computes the user acknowledge message:
UA = h(Bi^A || Nj || Nrc || SIDj), and sends UA to Sj.

A5. Sj computes UA' = h(Bi^A || Nj || Nrc || SIDj), and compares it with the received $UA$. As these are equal, Sj authenticates the adversary as a legitimate user, even though counterfeit credentials were used. The adversary can now start a session with Sj using the session key SK = h(Bi^A || Ni || Nj || Nrc || SIDj).

In this attack, the adversary can use random access credentials to impersonate a fictitious identity to the server without knowing either legitimate IDi or PWi. The root cause of the attack is that the adversary is capable of obtaining RC's secret $h(Krc)$ and then uses this secret to forge the essential value $Bi^A = h(N_{PWi}) \oplus h(Krc)$ during the protocol session. This forged value $Bi^A$ enables the creation of values DIDi and Qi. Meanwhile, the server only verifies that the same Bi component is used in the values DIDi and Qi. Hence, the server is unable to detect that fictitious credentials have been used, and it will accept the adversary's login request.

**3.2 New Attack against Hsiang and Shih Scheme**

In this section, we demonstrate that the Hsiang and Shih scheme suffers from a hitherto unknown weakness that an adversary can exploit to impersonate a fictitious user to access a server Sj.

*3.2.1 Hsiang and Shih Scheme.*
The Hsiang and Shih scheme has the same four phases as the Liao and Wang scheme. Initially, the registration centre RC creates Nrc, Krc and distributes h(SIDj || Nrc) to all registered servers. In the registration phase user Ui submits secret password h(Nb ⊕ PWi) and identity IDi to RC, who returns smart card containing (Vi, Bi, Hi, Ri, h(.)), where Ti = h(IDi || Krc); Vi = Ti ⊕ h(IDi || h(Nb ⊕ PWi)); Bi = Ai ⊕ h(Nb ⊕ PWi); Ri = h(h(Nb ⊕ PWi) || Nr); Hi = h(Ti); and Ai = h(h(Nb ⊕ PWi) || Nr) ⊕ h(Krc || Nr). The user Ui enters Nb into the smart card and the Login and Authentication & Verification phases authenticate the user to the server with the following steps, where DIDi = h(Nb ⊕ PWi) ⊕ h(Ti || Ai || Ni); Pij = Ti ⊕ h(Ai || Ni || SIDj); Qi = h(Bi || Ai || Ni); Di = Ri ⊕ SIDj ⊕ Ni; Co = h(Ai || Ni+1 || SIDj); Mjr = h(SIDj || Nrc) ⊕ Njr; C1 = h(Njr || h(SIDj || Nrc) || Nrj); C2 = Ai ⊕ h(h(SIDj || Nrc) ⊕ Njr); SA = h(Bi || Ni || Ai || SIDj); UA = h(Bi || Nj || Ai || SIDj); and  SK = h(Bi || Ai || Ni || Nj || SIDj).

1. Login Request U->S: (DIDi, Pij, Qi, Di, Co, Ni)
2. Login Server S->RC: (Mjr, SIDj, Di, Co, Ni)
3. RC Acknowledge RC->S: (C1, C2, Nrj)
4. Server Acknowledge S->U: (SA, Nj)
5. User Acknowledge U->S: (UA)
6. Message Exchange Using Session Key SK

The password change phase allows users to change their personal password.

*3.2.2 New Attack against Hsiang and Shih Scheme to Impersonate Fictitious Users.*
In addition to the known weaknesses discussed in section 2, we show that the Hsiang and Shih scheme suffers from the hitherto unknown weakness that can be exploited by impersonating a fictitious user to server *Sj*. The server is not able to detect that an illegal request has been made and will accept the login request as a legitimate.

In this scenario, the adversary is a legitimate user Ua with genuine access credentials from the registration centre RC. The adversary retrieves the secret tokens (Va, Ba, Ha, Ra, Nba, h(.)) from his/her own smart card and obtains $h(Krc \oplus Nr) = Ba \oplus h(Nba \oplus PWa) \oplus Ra$. Then the attack is mounted as follows:

A1. The adversary generates the random nonces: $N_{Ri}$ to replace Ri; $N_{SPWi}$ to replace h(Nb⊕PWi);  and $N_{Ti}$ to replace Ti. Also, the adversary computes the fictitious values: $A^A_i = N_{Ri} \oplus h(Krc \oplus Nr)$ and $B^A_i = A^A_i \oplus N_{SPWi}$.

A2. The adversary generates the nonce Ni and uses the obtained values to compute:

DIDi = h(Nb⊕PWi) ⊕ h(Ti || Ai || Ni) = $N_{SPWi}$ ⊕ h($N_{Ti}$ || $A^A_i$ || Ni)

Pij = Ti ⊕ h(Ai || Ni || SIDj) = $N_{Ti}$ ⊕ h($A^A_i$ || Ni || SIDj)

$Qi = h(Bi \parallel Ai \parallel Ni) = Qi = h(B^{A_i} \parallel A^{A_i} \parallel Ni)$

$Di = Ri \oplus SIDj \oplus Ni = N_{Ri} \oplus SIDj \oplus Ni$

$Co = h(Ai \parallel N_{i+1} \parallel SIDj) = h(A^{A_i} \parallel N_{i+1} \parallel SIDj)$

Subsequently, the adversary sends counterfeit login request (DIDi, Pij, Qi, Di, Co, Ni) to service provider Sj.

A3. Sj generates nonce Njr, computes $Mjr = h(SIDj \parallel Nrc) \oplus Njr$ and sends (Mjr, SIDj, Di, Co, Ni) to the registration centre RC.

A4. RC performs the following calculations:

$Njr = Mjr \oplus h(SIDj \parallel Nrc); Ri = Di \oplus SIDj \oplus Ni = N_{Ri};$

$Ai = Ri \oplus h(Krc \oplus Nr) = N_{Ri} \oplus h(Krc \oplus Nr) = A^{A_i};$ and $C'o = h(Ai \parallel N_{i+1} \parallel SIDj) = h(A^{A_i} \parallel N_{i+1} \parallel SIDj)$

The calculated value Co is compared with the received value Co*. As these match, RC accepts *Sj* as the legitimate server requested by the user. RC then generates nonce Nrj and computes:

$C1 = h(Njr \parallel h(SIDj \parallel Nrc) \parallel Nrj)$ and $C2 = Ai \oplus h(h(SIDj \parallel Nrc) \oplus Njr) = A^{A_i} \oplus h(h(SIDj \parallel Nrc) \oplus Njr)$

Finally, RC sends the RC Acknowledge Message (C1, C2, Nrj) to Sj.

A5. Sj computes $C1' = h(Njr \parallel h(SIDj \parallel Nrc) \parallel Nrj)$ and compares it with the received C1*. As they are equal, Sj authenticates RC and computes:

$Ai = C2 \oplus h(h(SIDj \parallel Nrc) \oplus Njr) = AA_i; Ti = Pij \oplus h(Ai \parallel Ni \parallel SIDj) = N_{Ti};$

$h(Nb \oplus PWi) = DIDi \oplus h(Ti \parallel Ai \parallel Ni) = N_{SPWi}; Bi = Ai \oplus h(Nb \oplus PWi) = A^{A_i} \oplus N_{SPWi} = B^{A_i};$

$Qi' = h(Bi \parallel Ai \parallel Ni) = h(B^{A_i} \parallel A^{A_i} \parallel Ni)$.

The calculated value Qi' is compared with the received Qi* from the user login request message. As these are equal, Sj generates nonce Nj, computes $SA = h(Bi \parallel Ni \parallel Ai \parallel SIDj)$ and sends (SA, Nj) back to the user.

A6. The adversary accepts the server acknowledgement message, computes the user acknowledge message $UA = h(B^{A_i} \parallel Nj \parallel A^{A_i} \parallel SIDj)$ and sends UA to Sj.

A7. Sj computes $UA' = h(B^{A_i} \parallel Nj \parallel A^{A_i} \parallel SIDj)$ and compares it with the received UA*. As these are equal, Sj authenticates the adversary as a legitimate user, even though counterfeit credentials were used. The adversary can now start a session with Sj using the session key $SK = h(B^{A_i} \parallel A^{A_i} \parallel Ni \parallel Nj \parallel SIDj)$.

In line with the attack on the Liao and Wang scheme, the adversary can use random access credentials to impersonate a fictitious identity to the server without knowing either legitimate IDi or PWi. The root cause of the attack is that the adversary can obtain RC's secret $h(Krc \oplus Nr)$ and uses this secret (together with random values $N_{Ri}$, $N_{SPWi}$, $N_{Ti}$) to forge the essential values $A^{A_i}$, $B^{A_i}$ during the protocol session. Meanwhile, RC will only establish whether the values Ai, Di use the same component $Ri$. This is achieved by retrieving $Ri$ from the received $Di$, using the retrieved $Ri$ value to compute values of Ai, Co' and comparing the computed Co' with the received value Co*. As these values match, RC is unable to detect that fictitious credentials have been used. RC will send the RC Acknowledge Message, which contains the forged $A^{A_i}$, to server Sj. As the server only verifies that the Login Request and the RC Acknowledge Message contain the same $Ai$, it is unable to detect that fictitious credentials have been used and will accept the bogus login request.

### 3.3 New Attacks against Lee et al. Scheme

In this section, we show that the Lee et al. scheme suffers from a hitherto unknown weakness that an adversary can exploit to impersonate a fictitious user to access a server *Sj.*

*3.3.1 Lee et al. Scheme.*

The Lee et al scheme has the same four phases as the Liao and Wang scheme. Initially, the registration centre RC creates $Nrc, Krc$ and distributes $h(Nrc), h(Krc \| Nrc)$ to all registered servers. In the registration phase user $U_i$ submits secret password $h(Nb \oplus PW_i)$ and identity $ID_i$ to RC, who returns smart card containing $(V_i, B_i, H_i, h(.), h(Nrc))$, where $T_i = h(ID_i \| Krc)$, $V_i = T_i \oplus h(ID_i \| h(Nb \oplus PW_i))$, $B_i = h(h(Nb \oplus PW_i) \| h(Krc \| Nrc))$ and $H_i = h(T_i)$. On receipt of the smart card, $U_i$ also stores $Nb$ in the smart card. The Login and Authentication & Verification phases authenticate the user to the server with the following steps, where $DID_i = h(Nb \oplus PW_i) \oplus h(T_i \| A_i \| N_i)$; $P_{ij} = T_i \oplus h(h(Nrc) \| N_i \| SID_j)$; $Q_i = h(B_i \| A_i \| N_i)$; $SA = h(B_i \| N_i \| A_i \| SID_j)$; $UA = h(B_i \| N_j \| A_i \| SID_j)$; $A_i = h(T_i \| h(Nrc) \| N_i)$; and $SK = h(B_i \| N_i \| N_j \| A_i \| SID_j)$:

1. Login Request          U->S: (DIDi, Pij, Qi, Ni)
2. Server Acknowledge     S->U: (SA, Nj)
3. User Acknowledge      U->S: (UA)
4. Message Exchange Using Session Key $SK$

The password change phase allows users to change their personal password.

*3.3.2 New Attack against Lee et al. Scheme to Impersonate Fictitious Users.*

In addition to the known weaknesses discussed in section 2, the Lee et al. scheme has an unknown weakness that can be exploited by an adversary to impersonate a fictitious user to access a server $S_j$. The server is not able to detect that an illegal request has been made and will accept the login request as a legitimate request.

In this scenario, the adversary is a legitimate user $U_a$ with genuine access credentials from the registration centre RC. The adversary retrieves the secret tokens $(V_a, B_a, H_a, R_a, Nb_a, h(.), h(Nrc))$ from his/her own smart card. Then the attack is mounted as follows:

A1. The adversary generates the random nonce $N_{Ti}$ to replace original value $T_i$ and uses the original values $PW_a$, $Nb_a$ and $B_a$ associated with the legitimate account $U_a$. It is worth noting that $PW_a$, $Nb_a$ and $B_a$ do not depend on the adversary's real identity $ID_a$ ($PW_i$ and $Nb$ can be changed by users at any time) and therefore, these cannot be used to reliably identify user $U_a$.

A2. As per protocol description, the adversary generates nonce $N_i$ and uses the obtained values to compute:

$A^{A_i} = h(T_i \| h(Nrc) \| N_i) = h(N_{Ti} \| h(Nrc) \| N_i)$; $DID_i = h(Nb \oplus PW_i) \oplus h(T_i \| A_i \| N_i) = h(Nb_a \oplus PW_a) \oplus h(N_{Ti} \| A^{A_i} \| N_i)$; $P_{ij} = T_i \oplus h(h(Nrc) \| N_i \| SID_j) = N_{Ti} \oplus h(h(Nrc) \| N_i \| SID_j)$; and $Q_i = h(B_i \| A_i \| N_i) = Q_i = h(B_a \| A^{A_i} \| N_i)$

Subsequently, the adversary sends the counterfeit login request $(DID_i, P_{ij}, Q_i, N_i)$ to the service provider $S_j$.

A3. $S_j$ attempts to validate the request and computes: $T_i = P_{ij} \oplus h(h(Nrc) \| N_i \| SID_j) = N_{Ti}$

The server $S_j$ is unable to detect that $T_i = NT_i$ rather than $T_i = h(ID_i \| Krc)$. Thus, $S_j$ uses this value to calculate: $A_i = h(T_i \| h(Nrc) \| N_i) = h(N_{Ti} \| h(Nrc) \| N_i)$; $h(Nb \oplus PW_i) = DID_i \oplus h(T_i \| A_i \| N_i) = h(Nb_a \oplus PW_a)$; $B_i = h(h(Nb \oplus PW_i) \| h(Krc \| Nrc)) = h(h(Nb_a \oplus PW_a) \| h(Krc \| Nrc)) = B_a$; and $Q_i' = h(B_i \| A_i \| N_i) = h(B_a \| A_i \| N_i)$. The calculated value $Q_i'$ is compared with the received $Q_i^*$ from the user login request. As these match, $S_j$ accepts the login request, computes $SA = h(B_i \| N_i \| A_i \| SID_j)$, and sends $(SA, N_j)$.

A4. The adversary accepts the server acknowledgement message, computes the user acknowledge message $UA = h(B_a \| N_j \| A_i \| SID_j)$ and sends UA to $S_j$.

A5. Sj computes UA' = h(Ba|| Nj || A$_i$ || SIDj) and compares it with the received UA*. As these are equal, Sj authenticates the adversary as a legitimate user, even though counterfeit credentials were used. The adversary can now start a session with Sj using the session key SK = h(Ba || Ni || Nj || Ai || SIDj).

In this attack, the adversary can use random access credentials to impersonate a fictitious identity to the server without knowing either legitimate IDi or PWi. The root cause of the attack is that this scheme misses dependencies in secrets: Values Ti, PWi, Nbi and Bi are the main components the server employs to authenticate users. However, there is no dependency between Ti (the user's secrete ID) and the other values. Thus, the adversary Ua can replace Ti with a random value N$_{Ti}$ and use it together with original values PWa, Nba and Ba associated with the legitimate account Ua to forge the login credentials. As neither of PWa, Nba and Ba is linked to the identity of the user account, the use of a random Ti value cannot be detected.

**3.4 New Attacks against Li et al. Scheme**

In this section, we show that the Li et al. scheme suffers from two hitherto unknown weaknesses can be exploited to access a server *Sj* without knowing *IDi* and *PWi*. We present an attack in section 3.4.2 where the adversary exploits a weakness to impersonate a fictitious user. In section 3.4.3, we present another new attack, where an adversary can use a stolen smart card to impersonate the owner of the stolen smart card.

*3.4.1 Li et al. Scheme.*

The Li et al. scheme has the same four phases as the Liao and Wang scheme
Initially, the registration centre RC creates Nrc, Krc and distributes h(Krc || Nrc) and h(SIDj || Nrc) to all registered servers. In the registration phase user Ui submits secret password Ai = h(Nb $\oplus$ PWi) and identity IDi to RC, who returns smart card containing (Ci, Di, Ei, h(.), h(Nrc)), where Bi = h(IDi || Krc), Ci = h(IDi || h(Nrc) || Ai), Di = h(Bi || h(Krc || Nrc)) and Ei = Bi $\oplus$ h(Krc || Nrc). On receipt of the smart card, Ui also stores Nb in the smart card. The Login and Authentication & Verification phases authenticate the user to the server with the following steps, where DIDi = Ai $\oplus$ h(Di || SIDj || Ni); Pij = Ei $\oplus$ h(h(SIDj || h(Nrc)) ||Ni ); M$_1$ = h(Pij || DIDi || Di || Ni); M$_2$ =h(SIDj|| h(Nrc)) $\oplus$ Ni; M$_3$ = h(Di || Ai || Nj || SIDj); M$_4$ =Ai $\oplus$ Ni $\oplus$ Nj; UA = h(Di || Ai || Ni || SIDj); and SK = h(Di||Ai|| Ni || Nj || SIDj):

1. Login Request          U->S: (DIDi, Pij, M$_1$, M$_2$)
2. Server Acknowledge    S->U: (M$_3$, M$_4$)
3. User Acknowledge      U->S: (UA)
4. Message Exchange Using Session Key SK

The password change phase allows users to change their personal password.

*3.4.2 New Attack against Li et al. Scheme to Impersonate Fictitious Users.*

In this scenario, the adversary has obtained a smart card through malicious means and retrieves the secret values (Ci, Di, Ei, Nb, h(.), h(Nrc)), from the stolen smart card. Then the attack is mounted as follows:
A1. The adversary generates the nonce N$_{Ai}$ to replace original token Ai and uses the retrieved tokens Di, and Ei associated with the legitimate owner of the smart card.
A2. As per protocol description, the adversary generates nonce Ni and uses the obtained values to compute:
    DIDi = Ai $\oplus$ h(Di || SIDj || Ni) = N$_{Ai}$ $\oplus$ h(Di || SIDj || Ni); Pij = Ei $\oplus$ h(h(SIDj || h(Nrc)) ||Ni );
    M$_1$ = h (Pij || DIDi || Di || Ni); and M$_2$ =h (SIDj|| h(Nrc)) $\oplus$ Ni.
Subsequently, the adversary sends counterfeit login request (DIDi, Pij, M$_1$, M$_2$) to service provider Sj.

A3. $S_j$ attempts to validate the request and computes:

$N_i = M_2 \oplus h(SID_j || h(N_{rc}))$; $E_i = P_{ij} \oplus h(h(SID_j || h(N_{rc})) || N_i)$; $B_i = E_i \oplus h(K_{rc} || N_{rc})$;

$D_i = h(B_i || h(K_{rc} || N_{rc}))$; $A_i = DID_i \oplus h(D_i || SID_j || N_i) = N_{Ai}$; and $M_1' = h(P_{ij} || DID_i || D_i || N_i)$.

The calculated value $M_1'$ is compared with the received $M_1^*$ from the login request. As these are equal, $S_j$ accepts the login request, generates nonce $N_j$ and computes the server acknowledgement message $(M_3, M_4)$: $M_3 = h(D_i || A_i || N_j || SID_j) = h(D_i || N_{Ai} || N_j || SID_j)$; and $M_4 = A_i \oplus N_i \oplus N_j = N_{Ai} \oplus N_i \oplus N_j$. $S_j$ responds with $(M_3, M_4)$ to the adversary.

A4. The adversary accepts the server acknowledgement message and computes:

$UA = h(D_i || A_i || N_i || SID_j) = h(D_i || N_{Ai} || N_i || SID_j)$. The adversary send $UA$ to $S_j$.

A5. $S_j$ computes $UA' = h(D_i || N_{Ai} || N_i || SID_j))$ and compares it with the received $UA^*$. As these are equal, $S_j$ authenticates the adversary as a legitimate user, even though counterfeit credentials were used. The adversary can now start a session with $S_j$ using session key $SK = h(D_i || A_i || N_i || N_j || SID_j)$.

In this attack, the adversary can use random access credentials to impersonate a fictitious identity to the server without knowing either legitimate $ID_i$ or $PW_i$. The root cause of the attack is that this scheme misses dependencies in secrets: Values $A_i$, $B_i$ and $D_i$ are the main components the server employs to authenticate users. However, there is no dependency between $A_i$ and the other values. Thus, the adversary $U_a$ can replace $A_i$ with a random value $NA_i$ and use it together with original values $B_i$ and $D_i$ associated with the victim's account $U_i$ to forge the login credentials. As the secret $A_i$ is not linked to the identity of the user account, the use of a random $A_i$ value cannot be detected. In summary, the adversary can use forged credentials to gain access to server $S_j$. Note: If the server $S_j$ keeps detailed records of the session, $S_j$ can contact the registration centre and use either the value $D_i$ or $E_i$ to track down the owner $U_i$ of the stolen smart card (both of which have $B_i$ as a component, which in turn contains $ID_i$). However, the attack cannot be traced back to the adversary.

*3.4.3 New Attack against Li et al. Scheme to Impersonate Owner of a Stolen Smart Card*

In this attack, an adversary can fool a service provider $S_j$ to authenticate the adversary as the legitimate owner $U_i$ of the smart card, even though the adversary does not know the corresponding personal password $PW_i$. In this scenario, the adversary has obtained the smart card of user $U_i$ through malicious means.

The adversary retrieves the secret values $(C_i, D_i, E_i, N_b, h(.), h(N_{rc}))$ from the stolen smart card. Further, the attacker requires a record of a login request message $(DID_{ik}, P_{ik}, M_{1k}, M_{2k})$ from the legitimate owner $U_i$ of the smart card to any service provider $S_k$. Then the attack is mounted as follows:

A1. The adversary retrieves $A_i$ from the recorded login message by computing:

$N_{ik} = M_{2k} \oplus h(SID_k || h(N_{rc}))$; and $A_i = DID_{ik} \oplus h(D_i || SID_k || N_{ik})$.

A2. As per protocol description, the adversary generates nonce $N_i$ and uses the obtained values to compute:

$P_{ij} = E_i \oplus h(h(SID_j || h(N_{rc})) || N_i)$; $DID_i = A_i \oplus h(D_i || SID_j || N_i) = N_{Ai} \oplus h(D_i || SID_j || N_i)$;

$M_1 = h(P_{ij} || DID_i || D_i || N_i)$; and $M_2 = h(SID_j || h(N_{rc})) \oplus N_i$.

With these values, the adversary can now construct the counterfeit login request message $(DID_i, P_{ij}, M_1, M_2)$, which is sent to service provider $S_j$. The remainder of this attack follows the original protocol specification.

The targeted service provider $S_j$ is unable to detect that the login message is counterfeit and will authenticate the adversary as user $U_i$. After the verification phase has finished, the adversary can communicate with $S_j$ using session key $SK = h(D_i || A_i || N_i || N_j || SID_j)$.

In this attack, the adversary can impersonate an owner of any stolen smart card to access servers without knowing either legitimate IDi or PWi. The root cause of this attack is that an attacker can extract usable tokens from a stolen smart card. These tokens can then be used to retrieve the value Ai from a previously recorded message. Thus, the adversary has all required components to authenticate as user Ui to any server Sj.

## 4 ANALYSIS AND PREVENTION OF NEWLY REVEALED WEAKNESSES ON DYNAMIC ID-BASED PASSWORD AUTHENTICATION SCHEMES

The new attacks presented in the previous section demonstrate that none of the examined schemes can guarantee mutual authentication, as an adversary can forge an apparently valid login request. The authentication schemes of Liao and Wang, Hsiang and Shih, Lee et al. and Li et al., all use a dynamic identity (value DIDi) to protect the user anonymity during the session with server Sj. As information can be extracted from smart cards [30], a user password PWi is used to provide two-factor security. The authors claim that their schemes are secure, as long as only the password or the smart card, but not both, are obtained by an adversary [10, 11, 12, 13]. Further, the design requirements of these schemes state that no verification or password tables should be maintained in the servers. Thus, the basic steps for authentication in these schemes are:

- Users calculate authentication tokens based on users' password, their real identity and information stored in the smart card.
- Servers verify the users' authentication tokens by comparing them against re-computed values obtained from the components received from users and system-wide information obtained from the registration centre during the server registration phase.

Thus, Sj essentially establishes whether the values received from $Ui$ are consistent with the values sent from a user who has registered with the registration centre RC. However, no attempt is made to establish whether the user is in fact registered with RC.

### 4.1 Investigation Results

Investigating the structure of the communication messages in all the discussed authentication schemes reveals that all these schemes fail to fulfil the following required conditions:

- C1. *Non-disclosure of Registration Centre secrets:* If an adversary can learn Registration Centre secrets, the adversary can obviously use these to create forged credentials that a server cannot distinguish from legitimate credentials. Registration Centre secrets include values such as $Krc, h(Krc), h(Krc \oplus Nr)$.
- C2. *Sufficient dependencies between user-submitted values*: Servers use components received from the user to re-compute authentication tokens. If there are insufficient dependencies between these user-submitted values, then an adversary that is also a legitimate user might be able to use a combination of random values and values obtained from the legitimate smart card to forge authentication tokens.
- C3. *Inability of unauthorized parties to retrieve usable tokens from smart cards*: All authentication schemes discussed in this paper assume that an attacker can retrieve values from a smart card that has been obtained by malicious means. Thus, it is important to store authentication tokens in a protected manner that prevents an unauthorized party to use these values in a bogus authentication request.

In authentication schemes were condition C1 is violated, the adversary first retrieves secret RC information from a smart card and then uses the retrieved information to forge suitable credentials to impersonate a fictitious user. As the forged credentials are created using bona fide RC secrets, servers are not able to detect the forgery and will accept the forged credentials. For example, consider the attack on the Liao and Wang scheme detailed in Section 3.1.2. The adversary, which is a legitimate user of the system, uses his/her password PWi to retrieve $h(Krc) = Bi \oplus h(PWi)$ from the smart card. Subsequently, the attacker uses two random values $N_{Ti}$ and $N_{PWi}$. The former, $N_{Ti}$ is used to replace Ti and $N_{PWi}$ is used in conjunction with the retrieved $h(Krc)$ to calculate a fictitious Bi. These values are then used to forge a request to server Sj, who is not able to detect the request as being illegitimate. Thus, the leakage of RC's secret $h(Krc)$ leads to the fictitious user attack, where an adversary can impersonate a fictitious user to gain access to server.

Authentication schemes that violate condition C2, allow attackers to use a combination of random values together with legitimate user components. If there are insufficient interdependencies between these values, then servers are not able to detect that the random values are not related to the legitimate components and they might accept bogus authentication requests. For example, consider the attack against the Lee et al. scheme outlined in Section 3.3.2. The adversary is a legitimate user Ua that attempts to authenticate as a fictitious user. The adversary replaces the value Ti with a random value $N_{Ti}$ and uses the original values PWa, Nba and Ba associated with the legitimate account Ua. As there are no dependencies between Ti and PWi, Nbi, Bi, the computations performed by server Sj all yield the expected result. Thus, *Sj* is not able to detect that token $N_{Ti}$ is a counterfeit value, rather than value $h(IDi \| Krc)$, that corresponds to the received values derived from PWa, Nba and Ba. Therefore, the server accepts the counterfeit authentication request as a legitimate request.

If an authentication scheme violates condition C3, then the attacker can directly use values from a smart card that has been obtained by malicious means to mount the attack. For example, consider again the attack against the Li et al. scheme outlined in Section 3.4.2. In addition to violate condition C2 (sufficient dependencies between user-submitted values) it also violates condition C3: The adversary is able to retrieve values Di, Ei from the smart car (as these are stored in an unprotected manner) and can use these values in the attack. As these values are legal values created by RC, severs are unable to detect that they are used by an unauthorized entity.

In summary, the new attacks revealed in this paper succeed, as servers do not possess sufficient knowledge about users' access credentials. Servers re-compute these credentials based on received values and secrets shared with the registration centre and compare the computed values with the received values. If they match, servers assume successful authentication. In cases where the conditions C1 and C2 are not met, adversaries can construct suitable credentials based on secret values extracted from the smart card and random values that servers cannot distinguish from genuine credentials.

**4.2 Design Guidelines for Attack Prevention**

In this section, we propose a set of design guidelines for dynamic ID-based remote user authentication schemes in multi-server environments in order to prevent the weaknesses revealed during our investigation.

- DG1. *No passwords exposure:* The user's password should not be revealed to any principals, servers or the registration centre. If the user's password is revealed during registration or authentication, any administrator of the server/registration centre could use the identity and password to impersonate users.
- DG2. *Efficient password procedures:* The authentication scheme should feature efficient and convenient procedures for users to select and update their personal passwords. Further, the authentication mechanism should provide an efficient solution for dealing with incorrect passwords. In general, this can be achieved by implementing authentication as close to the user as possible, e.g. if a smart card is used, the user should be authenticated locally by the smart card without interaction with any server.

- DG3. *Safeguarding of RC secrets:* Implementation of this enhancement requires identification of all RC secrets used to compute tokens that are stored in smart cards or used in the login and verification phase and then ensuring that these secrets cannot be utilized by users to forge tokens.
- DG4. *Ensuring sufficient interdependencies between authentication tokens:* Authentication tokens submitted by users in the login request require sufficient interdependencies to ensure that the attackers cannot mix legitimate tokens with random values or other counterfeit tokens.
- DG5. *Protected stored authentication tokens:* If a possession factor such as a smart card is used, then any authentication tokens stored in the smart card should be stored in a protected manner that prevents extraction of usable tokens by unauthorized entities. As has been demonstrated in this work, if servers authenticate users only based on information directly stored in a smart card, then an adversary can retrieve these tokens from a maliciously obtained smart card to impersonate users.
- DG6. *Multi-factor authentication:* Authentication should be based upon at least two factors of the following categories 1) knowledge factors; 2) possession factors; and 3) inherence factors. For example, a password is a knowledge factor and a smart card is a possession factor. In such a scenario, a strong password should be used to authenticate the legitimate user to the smart card. On successful user authentication, the smart card will initiate a mutual authentication process with a server.
- DG7. *Mutual authentication:* Mutual authentication should be achieved between the user and the corresponding remote systems/server. The user can authenticate the identity of the server and the server can verify the identity of the user.
- DG8. *Session key agreement:* A session key should be established between user and server, once mutual authentication has been achieved. Any subsequent communication is encrypted with this session key, which provides confidentiality and secrecy of the transmitted data. In any further sessions a new key needs to be established. Both server and user need to be able to verify freshness of the agreed session key.
- DG9. *Forward secrecy:* Forward secrecy indicates that if the master secret key of any server is compromised, the secrecy of any previously established session keys should not be affected.
- DG10. *User anonymity:* User anonymity ensures that an eavesdropper cannot identify which user is involved in a communication with a server. Consequently, an adversary cannot analyze the activities being performed by a specific user.
- DG11. *Possession factor revocation:* If a possession factor (e.g. a smart card) is lost, users need the ability to revoke this possession factor. Thus, the system needs to provide a mechanism to invalidate a possession factor that ensures that it will not be accepted in any future authentication processes, even if an attacker should obtain all required authentication factors (e.g. an attacker may learn the users password through social engineering).
- DG12. *Resistance to attacks:* A secure protocol design needs to be resistant against various attacks such as forgery attacks [33]; replay attacks [3, 36]; stolen smart card attacks [10]; parallel session attacks [3, 31, 32]; stolen-verifier attack [34], insider attack [12, 37] and denial-of-service [9, 38].

### 4.3 Application of Proposed Guidelines

The conformance of the four discussed schemes [10, 11, 12, 13] is established by examining their compliance with the twelve above proposed guidelines. The results of this evaluation are summarized in Table 2. The second column of this table indicates the violated design guidelines, while the third column presents the root cause of the attack discovered on each scheme. This study shows that for all the schemes evaluated with new discovered attacks violate at least one of the guidelines. Hence, the application of the proposed guidelines shows how non-conformance with these guidelines causes the protocols to be attackable.

Table 2: Empirical results of proposed design guidelines

| Analyzed Protocol | Violated guidelines | The root cause of the attack |
|---|---|---|
| Liao and Wang Scheme | DG3, DG5 | Adversary can obtain the secret of RC: $h(Krc)$. |
| Hsiang and Shih Scheme | DG3, DG5 | Adversary can obtain the secret of RC: $h(Krc \oplus Nr)$. |
| Li et al. Scheme | DG4 | The scheme misses dependencies in secrets. |
| Li et al. Scheme | DG4, DG5 | The scheme misses dependencies in secrets and the adversary can extract usable tokens from a stolen smart card. |

## 5  CONCLUSION

In this paper, four dynamic ID-based password authentication schemes for multi-server environments were reviewed. Further, several hitherto unknown attacks against these schemes were revealed. The presented new attacks ascertain that these protocols fail to achieve mutual authentication and it was demonstrated how an adversary can impersonate either a fictitious or a legitimate user to any server. These attacks are possible, as the servers re-compute authentication tokens based on received values and secrets shared with the registration centre. By crafting suitable counterfeit credentials based on secret values extracted from the smart card, an adversary can trick a server to accept these counterfeit credentials.

We investigated the weaknesses exploitable by the discussed dynamic ID-based authentication schemes and established the reasons for the success of the new attacks. Furthermore, we proposed a new set of design guidelines to prevent such exploitable weaknesses on dynamic ID-based remote user authentication schemes in multi-server environments and we demonstrated how non-conformance with these guidelines causes the analyzed protocols to be attackable.